\documentclass[11pt]{article}
\usepackage[utf8]{inputenc}
\usepackage{url}
\usepackage{algorithm}
\usepackage[noend]{algorithmic}

\usepackage{amssymb,xy}
\usepackage{amsthm, amsmath}
\usepackage{graphicx}
\usepackage[vmargin=3cm,hmargin=2cm]{geometry}
\usepackage{fancyhdr}

\DeclareMathOperator{\distance}{d}
\newtheorem{theorem}{Theorem}

\begin{document}
\title{Minimum distance computation of linear codes via genetic algorithms with permutation encoding}

\author{
Jos\'e G\'omez-Torrecillas\({}^*\), F. J. Lobillo\({}^*\) and Gabriel Navarro${}^\dagger$\\
\\
\({}^*\) Department of Algebra and CITIC, University of Granada\\
${}^\dagger$ Department of Computer Sciences and AI, and CITIC, University of Granada
 \\
\url{gomezj@ugr.es, jlobillo@ugr.es, gnavarro@ugr.es}\\
}

\date{}

\maketitle

\begin{abstract}
We design a heuristic method, a genetic algorithm, for the computation of an upper bound of the  minimum distance of a linear code over a finite field. By the use of the row reduced echelon form, we obtain a permutation encoding of the problem, so that  its space of solutions does not depend on the size of the base field or the dimension of the code. Actually, the efficiency of our method only grows non-polynomially with respect to the length of the code.
\end{abstract}

\section{Introduction}

The minimum distance is a fundamental parameter to be computed when evaluating the practical utility of a linear code. This is so, since it allows to know its error-correcting capability.  It is well-known that this calculation is an NP-hard problem, and the associated decision problem is NP-complete  \cite{vardy}. Therefore, unless P=NP, it is a hopeless task to design an exact algorithm for finding the minimum distance of any code in a reasonable time. Among the developed algorithms, the fastest is the celebrated Brouwer-Zimmermann (BZ) algorithm, see \cite{wassermann}. Despite the BZ algorithm can be applied to codes over any finite field, in practice, it can be considered effective only for binary codes.  Nevertheless, recently, the use of large finite fields is being taken into consideration. For instance, the design of skew cyclic  codes \cite{GLN_IEEE}, both block or convolutional, and decoding algorithms \cite{GLN_IEEE2, GLN_PGZ}  require finite fields of large size, see examples in \cite{GLNN}. In the literature it can be found some approximate algorithms as, for instance, in \cite{askali} or \cite{leon}. Nevertheless, once again, the space of possible solutions grows exponentially with respect to the bit-size of the elements of the base field (i.e. the dimension of the base field over its prime subfield), as well as with respect to the dimension of the code. Our proposal consists of the design and implementation of an approximate algorithm whose space of solutions only depends on the length of the code, so that its efficiency grows polynomially with respect to the bit-size of the elements and the dimension. Concretely, we design a genetic algorithm, based on the generational model, for computing an upper bound of the distance.

\section{Permutation encoding of the problem}
\begin{figure}
$$\includegraphics[width=10cm]{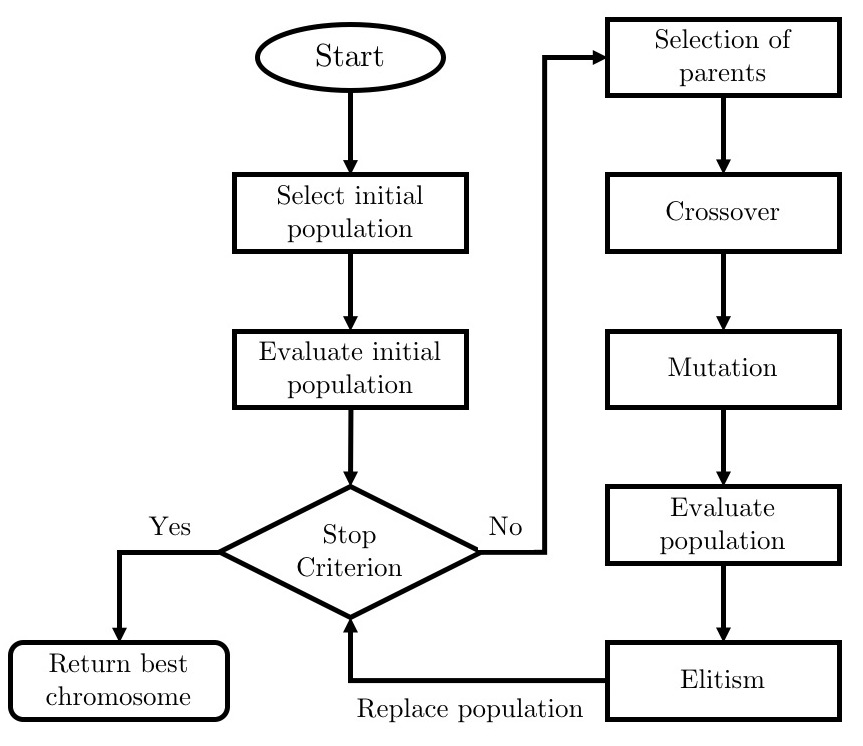}$$
\caption{Standard scheme of a generational genetic algorithm}\label{f1}
\end{figure}

Genetic algorithms are a celebrated class of heuristic methods that follows a biologically inspired search model in order to solve optimization problems. Concretely, from a population of possible solutions (\emph{chromosomes}), they simulate the process of genetic recombination, see \cite[Chapter 3]{meta} for a basic reference. Each chromosome has attached its image by the map under consideration, its \emph{fitness}, which measures its adaptation to the problem. By \emph{crossover} and \emph{mutation} operators, the population evolves so that, hopefully, a chromosome in it provides an optimum of the problem, i.e. its fitness reaches the optimum value.  In Figure \ref{f1}, we show the standard scheme of a generational genetic algorithm.

The development of a genetic algorithm requires first to establish an encoding of the space of solutions of the problem. For finding the minimum distance of an $[n,k]_q$ linear code $\mathcal{C}$, an obvious way is to consider $k$-tuples over the finite field $\mathbb{F}_q$, that represent the possible linear combinations of the rows of a generating matrix $G$ of $\mathcal{C}$. This is done  in \cite{askali} for binary codes. Nevertheless, this space of solutions grows exponentially with respect to the bit-size of the elements of the field and the dimension. In contrast, our proposal only depends on the length of the code. We shall need the following result. Its proof is not complicated, but, as far as we have searched, we have not found it in the literature.

\begin{theorem}
Let $G$ be a $k\times n$ generating matrix of a $[n,k]_q$-linear code $\mathcal{C}$ over the finite field $\mathbb{F}_q$. There exists a permutation $P\in\mathcal{S}_n$ such that the row reduced echelon form $R$ of $GM_P$, where $M_P$ is the permutation matrix of $P$, satisfies that the Hamming weight of some of its rows equal the minimum distance of $\mathcal{C}$. Consequently, if $b$ is a row of $R$ verifying such property, then $bM_P^{-1}$ is a codeword of minimal weight of $\mathcal{C}$ .
\end{theorem}
\begin{proof}
Let $d=\distance (\mathcal{C})$ be the minimum distance of \(\mathcal{C}\), then there exists a non singular matrix $A\in \mathcal{M}_k(\mathbb{F}_q)$ and a permutation $P\in \mathcal{S}_n$ such that
$$AGM_P=\left ( \begin{array}{ccccccc} & & &G_1& &  \\ \hline
0 & \cdots & 0 & | & a_1 & \cdots & a_d \end{array} \right ),$$
where $a_1,\ldots ,a_d$ are nonzero. Now, there exists an invertible matrix $A'\in \mathcal{M}_{k-1}(\mathbb{F}_q)$ such that  $AG_1$ is the row reduced matrix $R'$ of $G_1$. Hence,
\begin{equation}\label{rref}
\left (\begin{array}{c|c} A' & 0\\ \hline 0& 1 \end{array}  \right )AGM_P=\left ( \begin{array}{ccccccc} & & &R'& &  \\ \hline
0 & \cdots & 0 & | & a_1 & \cdots & a_d \end{array} \right ).
\end{equation}
Since $G_1$ has rank $k-1$, the last row of $R'$ is nonzero. So assume that the pivot of this row  is in the $i_0$-th column. If $i_0<n-d+1$, then the last row of \eqref{rref} is the last row of the row reduced echelon form of $GM_P$ up to non zero scalar multiplication, and we are done. Otherwise, the last two rows of \eqref{rref} are linearly independent and their nonzero coordinates are placed at the last $d$ coordinates. Hence, there exists a linear combination of both whose hamming weight is lower than $d$, a contradiction. The last statement is straightforward.
\end{proof}

\noindent Therefore, the problem is reduced to find the minimum of the map $\mathfrak{d}:\mathcal{S}_n\to \mathbb{N}$ defined by
\[
\mathfrak{d}(P)=\min\{\mathrm{w}(b) \, | \, b \text{ is a row of the row reduced echelon form of } GM_P\},
\]
the \emph{fitness} of the permutation $P$, where $\mathrm{w}(b)$ denotes the Hamming weight of $b$. This encoding is then invariant with respect to the base field. Obviously, the computation of $\mathfrak{d}(P)$, for some permutation $P$, does depend on $q$ and $k$. However, it can be calculated by $\mathcal{O}(k^2n)$ operations in $\mathbb{F}_q$.

\section{The genetic algorithm}

A genetic algorithm starts with an initial population of chromosomes that evolves. In our algorithm we follow the most common strategy and the initial population is selected randomly. The key point is then to decide how the population evolves by crossover and mutation operators. This has to be chosen appropriately in order to get a suitable balance between diversity when exploring the search space, and convergence in promising zones.  We first select randomly the chromosomes to be crossed with certain probability, say $p_c$. The classic crossover operators do not consider the group structure of $\mathcal{S}_n$. Intuitively, for a permutation, the more non-information columns it moves to the first $k$ positions, the better fitness it has. Therefore, one could expect that the composition of permutations with good fitness, may produce a chromosome with better fitness. Additionally, since two (or more) random permutations in $\mathcal{S}_n$ probably form a generator system \cite[Theorem 1]{dickson}, the whole space of solutions is reached by their composition. In this paper we propose to use the following family of algebraic crossovers: given $r$ chromosomes $\chi_1,\chi_2,\ldots ,\chi_r$, we construct $$\mathcal{T}=\{\chi_{\tau{(1)}}\circ \chi_{\tau{(2)}}\circ \cdots \circ \chi_{\tau{(r)}} \text{ such that } \tau\in \mathcal{S}_r\}.$$
From this set, we select the $r$ chromosomes with lower image under $\mathfrak{d}$ (that is, with better fitness) which replace the original $r$ chromosomes. Therefore, the algebraic crossover operator $AX_r$ partitions the population into subsets of $r$ elements and, for each subset, with a given probability $p_c$, it recombines the elements as described above.

\floatname{algorithm}{Procedure}
\begin{algorithm}[ht]
\caption{$\operatorname{NextGeneration}$}\label{RG}
\begin{algorithmic}[1]
\REQUIRE $P(i)$, population at time $i\geq 0$; $G$, generating matrix; $p_c$, crossover probability; $p_m$, mutation probability; $r$, number of chromosomes involved in the crossover step.
\ENSURE $P(i+1)$ population at time $i+1$.
\STATE $best \gets$ chromosome that reaches the minimum of  $\mathfrak{d}$ in $P(i)$.
\STATE $X\gets \emptyset$
\WHILE{$\#P(i)\geq r$}
\STATE $\mathcal{S}\gets$ $\{ r \text{ randomly chosen chromosomes of } P(i)\}$.
\STATE $P(i)\gets P(i)-\mathcal{S}$
\STATE $\mathcal{S}\gets AX_r(\mathcal{S})$ with probability $p_c$.
\IF{$\mathcal{S}$ was not crossed}
\FOR{$s\in \mathcal{S}$}
\STATE $s\gets s\circ t$ for a random transposition $t=(t_1,t_2)$, where $t_1\leq k$ and $t_2>k$ ,with prob. $p_m$.
\ENDFOR
\ENDIF
\STATE $X\gets X\cup \mathcal{S}$
\ENDWHILE
\STATE $X\gets X\cup P(i)$
\STATE Change the worst of $X$ by $best$, if $best$ is not in $X$.
\RETURN \( X\)
\end{algorithmic}
\end{algorithm}

In the mutation step, we shall follow a standard mutation operator: the composition with a transposition. Nevertheless, permuting two ``non-pivot'' columns does not modify the fitness. Therefore we wanted to force to choose randomly a ``pivot'' column and a ``non-pivot'' column.  For reasons of efficiency, we simply choose a column from the first $k$ columns and other from the remaining $n-k$ columns, where $k$ is the dimension of the code. The mutation operator will be then applied, with probability $p_m$, to those chromosomes that were not crossed.

Finally, in order to ensure convergence to the optimum, we add the best chromosome of the older generation to the new one (if it was not). Procedure \ref{RG} comprises the computation of a new generation of chromosomes, whilst Algorithm \ref{GAGLN} describe the whole genetic algorithm.

\floatname{algorithm}{Algorithm}
\begin{algorithm}
\caption{Genetic algorithm for distance aproximation}\label{GAGLN}
\begin{algorithmic}[1]
\REQUIRE $(G, r, p_m, p_c)$, as in Procedure \ref{RG}; $c$ or $t$, number of iterations or execution time, respectively; $p$, size of the population.
\ENSURE $\overline{d}$ upper bound of the minimum distance of $\mathcal{C}$.
\STATE $P(0)\gets$ random initial population of size $p$.
\STATE $i\gets 1$
\WHILE{$i<c$ (or $time<t$)}
\STATE $P(i)\gets \operatorname{NextGeneration}(P(i-1),G,r, p_c,p_m)$
\STATE $i\gets i+1$
\ENDWHILE
\STATE $best \gets$ chromosome that reaches the minimum of  $\mathfrak{d}$ in $P(i)$.
\RETURN \(\mathfrak{d}(best)\)
\end{algorithmic}
\end{algorithm}

%%%
\section{A small example}

\begin{figure}
$$\includegraphics[width=13cm]{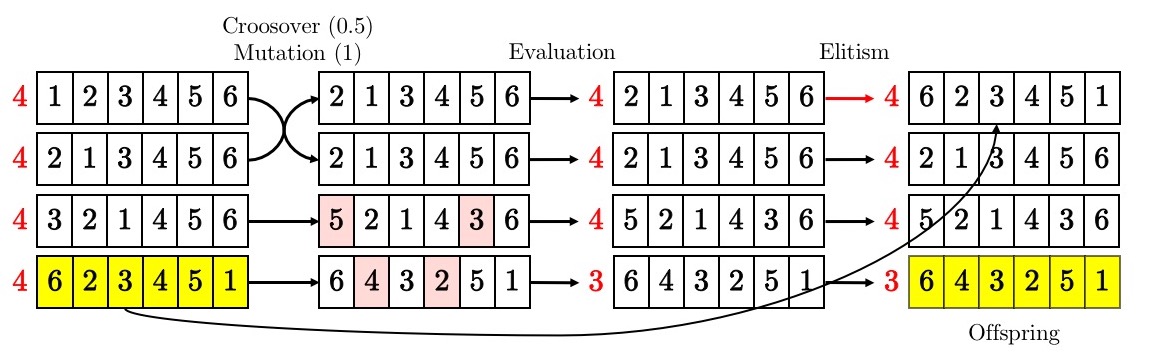}$$
\caption{Construction of the next generation}\label{f2}
\end{figure}

Let $\mathcal{C}$ be the $[6,3]$-linear code over $\mathbb{F}=\mathbb{F}_8=\mathbb{F}_2(a)$ with generating matrix $$G=   	
\left(\begin{array}{rrrrrr}
a^{5} & 0 & a^{5} & a^{6}  & a &
0 \\
a^{4}  & a & 1 & 0 & a & a^{2} \\
a^{5}  & a^{4} & a^{6} & a^{4} &
a^{2} & 1
\end{array}\right)\in \mathcal{M}_{3\times 6}(\mathbb{F}).$$ 
Suppose that we start with an initial population of 4 chromosomes, that we evaluate. We have marked with yellow color the best chromosome of the population.
$$\includegraphics[width=3cm]{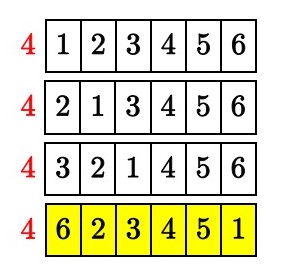}$$
Hence, for $AX_2$, $p_c=0.5$ and $p_m=1$, the execution of Procedure  \ref{RG} is described in Figure \ref{f2}.
\section{Experiments}\label{exp}
We show here a little experiment of the performance of Algorithm \ref{GAGLN}. It was run by an ad-hoc implementation in C++.  The executions have been done by a processor Intel Core i7 3GHz under macOS 10.12.6. Nevertheless, in order to avoid the run-time dependency from the chosen programming language and processor, we show the number of times that the population has been evolved. We consider some linear codes over $\mathbb{F}_8$ of the database \url{http://codetables.de} in Table \ref{t1}. We want to point out that an implementation of the BZ Algorithm for the $[30,14,12]_8$-linear code is estimated to take more than 20 hours of execution. The QR code of length 223 is studied in \cite{saouter} in order to check that its distance is 31.
\begin{table}[ht]
\centering
\begin{tabular}{cccc||ccccc}
\hline
field & length & dim.  & dist.  & dist. approx. & pop. size & loop &  time (sec.) & random \\  \hline \hline
$\mathbb{F}_8$&30 & 14 & 12 & \textbf{12}\tiny{(100)} & 5 & 0 & $\sim 0$ & \textbf{12}\tiny{(100)} \\
%$\mathbb{F}_8$&45 & 24 & 14 & \textbf{14} & & 1 & $10^{-6}$ &14 \\
$\mathbb{F}_8$&60 & 30 & 20 & \textbf{20}\tiny{(100)} & 5 & 0 & $ \sim 0$ & \textbf{20}\tiny{(100)}  \\
%$\mathbb{F}_8$&75 & 45 & 17 & \textbf{17} &  & 3 & 0.3  & \\
$\mathbb{F}_8$&90 & 19 & 49 &  \textbf{49}\tiny{(100)} & 10 & 61 & 0.03 & \textbf{49}\tiny{(96)} \\
$\mathbb{F}_8$&90 & 50 & 21 & 22\tiny{(49)} &20 & 423 &  1.69 & 22\tiny{(3)}  \\
$\mathbb{F}_8$&90 & 60 & 16 & \textbf{16}\tiny{(94)} & 30 & 284 &  1.91 & \textbf{16}\tiny{(37)} \\
$\mathbb{F}_8$&130 & 75 & 28 & \textbf{28}\tiny{(10)} & 150 & 524 & 40.35  & 30\tiny{(4)} \\
$\mathbb{F}_8$&130 &  85 & 23 & \textbf{23}\tiny{(3)} & 150 & 471 & 40.11  &  24\tiny{(12)}\\
$\mathbb{F}_8$&130 & 95 & 18 & \textbf{18}\tiny{(91)} & 50 & 320 & 9.58 & \textbf{18}\tiny{(16)} \\
$\mathbb{F}_2$&QR(223) & 112 & 31 & \textbf{31}\tiny{(100)} & 5 & 39 &  0.7  & \textbf{31}\tiny{(100)} \\ \hline
\end{tabular}
\caption{Execution of Algorithm \ref{GAGLN} for $AX_2$, $p_c=0.7$, $p_m=1$, where the populations evolve a maximum of 1000 generations. 100 repetitions for each code. In the distance approximations, between parenthesis, the number of times that the best bound is reached. The columns ``loop'' and ``time'' show the average number of generations and the mean time for which the best bound is reached. The column ``random'' shows the lowest weight obtained by selecting randomly $1000p$ permutations, where $p$ is the size of the population.}\label{t1}
\end{table}

\begin{figure}
$$\includegraphics[width=6cm]{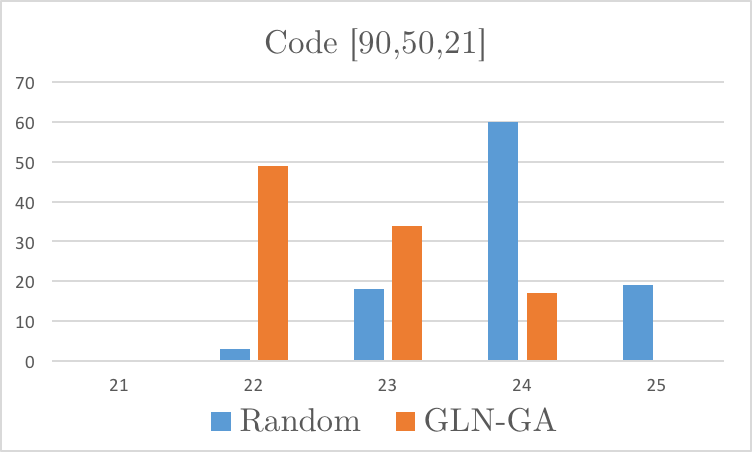}  \quad \includegraphics[width=6cm]{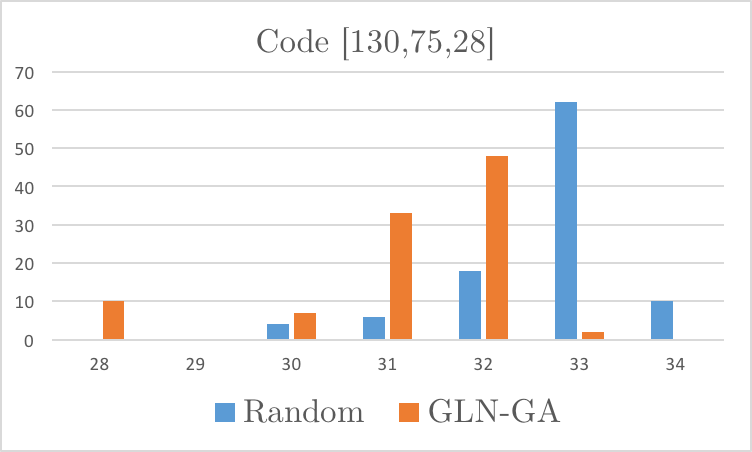}$$
$$\includegraphics[width=6cm]{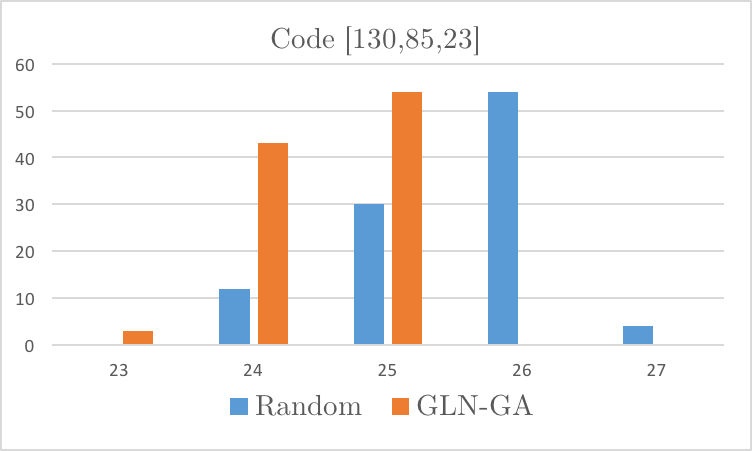}\quad \includegraphics[width=6cm]{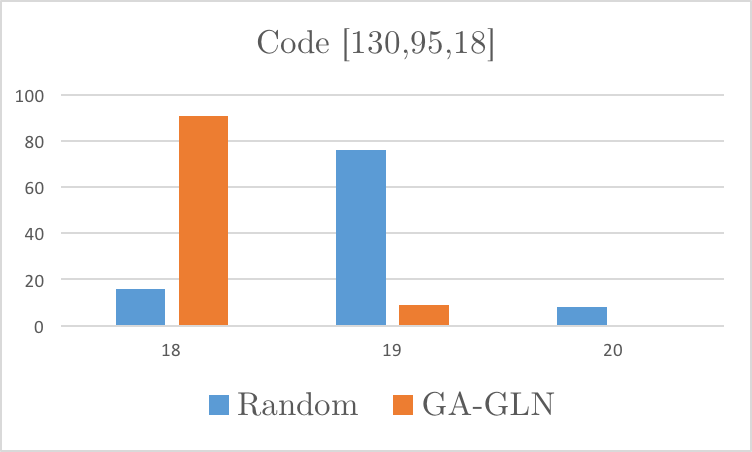}$$
\caption{Algorithm \ref{GAGLN} vs randomness}\label{f3}
\end{figure}

Additionally, in Figure \ref{f3}, we show the distributions of the distances obtained for some codes of Table \ref{t1} for Algorithm \ref{GAGLN} and the random selection of 1000 generations.

\section{Conclusion}
This paper comprises a first approach to the computation of the minimum distance of linear codes over large fields by heuristic methods. Due to the nature of the problem, the resolution by exact algorithms seems to be hopeless. So, our proposal considers the application of genetic algorithms with permutation encoding, which eliminates the exponential dependency on the bit-size of the elements of the base field. Future improvements should take into account the refinement of the space of solutions or the design of good performance metaheuristics for permutation encodings, as, for instance, ant colony optimization.

\begin{quote}
{\bf This research has been supported by grant MTM2016-78364-P from the Spanish Agencia Estatal de Investigaci\'on and FEDER.}
\end{quote}

\end{document}